\documentstyle[aps]{revtex}

\begin{document}

\preprint{HIP-1999-84/TH}

\title{Higgs sector and R-parity breaking couplings in models with
broken $U(1)_{B-L}$ gauge symmetry}

\author{Kai Puolam{\"{a}}ki}

\address{Helsinki Institute of Physics, P.O.Box 9,
FIN-00014 University of Helsinki, Finland}

\date{April 27, 2000\footnote{Original version submitted to Physical
Review D on December 30, 1999.}}

\maketitle

\begin{abstract}

Four different supersymmetric models based on $SU(2)_L \times U(1)_R
\times U(1)_{B-L}$ and $SU(2)_L \times SU(2)_R \times U(1)_{B-L}$
gauge symmetry groups are studied. $U(1)_{B-L}$ symmetry is broken
spontaneously by a vacuum expectation value (VEV) of a sneutrino
field. The right-handed gauge bosons may obtain their mass solely by
sneutrino VEV. The physical charged lepton and neutrino are mixtures
of gauginos, higgsinos and lepton interaction eigenstates. Explicit
formulae for masses and mixings in the physical lepton fields are
found. The spontaneous symmetry breaking mechanism fixes the trilinear
R-parity breaking couplings. Only some special R-parity breaking
trilinear couplings are allowed. There is a potentially large
trilinear lepton number breaking coupling --- which is unique to
left-right models --- that is proportional to the $SU(2)_R$ gauge
coupling $g_R$. The couplings are parametrized by few mixing angles,
making the spontaneous R-parity breaking a natural ``unification
framework'' for R-parity breaking couplings in SUSYLR models.

\end{abstract}

%See <URL: http://www.aip.org/pacs/> for PACS numbers.
\pacs{11.30.Pb, 11.30.Qc, 12.60.Jv}
% 11.30.Pb       Supersymmetry (see also 12.60.J Supersymmetric
% models)
% 12.10.Dm       Unified theories and models of strong and electroweak
%                interactions
% 11.30.Qc       Spontaneous and radiative symmetry breaking
% 12.60.Jv       Supersymmetric models (see also 04.65 Supergravity)

\section{Introduction}

A major problem in supersymmetry is related to the lepton and baryon
numbers, which seem to be conserved to a very high precision. In the
standard model (SM) lepton- or baryon number violating renormalizable
interactions do not exist due to the particle content and
gauge symmetry. In the minimal supersymmetric standard model (MSSM),
instead, given all the supersymmetric partners of standard model
particles, one would expect {\em a priori} both lepton and baryon
number to be violated. On the baryon and lepton number violating
couplings there are, however, strong experimental constraints. The
most notable of the limits follows from the non-observation of nucleon
decay, which sets extremely stringent limits on the products of lepton
and baryon number violating couplings \cite{nucleondecay}.

One can cure the problem by assuming that so-called R-parity is
conserved. R-parity is defined by $R=(-1)^{3(B-L)+2S}$ where $B$ and
$L$ are the baryon and lepton numbers of respective fields and $S$ is
spin. If R-parity is conserved the proton is stable. Also the lightest
supersymmetric particle (LSP), which usually is neutralino, does not
decay and is thus a good candidate for dark matter. Due to conserved
R-parity the supersymmetric particles can be only produced in pairs in
collider experiments~\cite{rparity}. Conservation of R-parity is a
much stronger assumption than is phenomenologically necessary. It
suffices that either baryon or lepton number violating interactions
are strongly suppressed to avoid proton decay, and that the remaining
interactions are small enough not to have been directly observed.

If the R-parity were a gauge symmetry, it would be protected against
violations arising for example from quantum gravity. Attractive
alternative to a global symmetry would thus be a local R-parity. This
can be realized in a theory based on a gauge group that has $B-L$
symmetry as a discrete subgroup. An interesting low energy theory with
this property is the supersymmetric left-right (SUSYLR) theory obeying
the gauge symmetry $SU(3)_C \times SU(2)_L \times SU(2)_R \times
U(1)_{B-L}$, where the R-parity is a discrete subgroup of $U(1)_{B-L}$
gauge symmetry. This model can be embedded in a supersymmetric
$SO(10)$ theory~\cite{so10gut,model2bnor}.

It is possible that in the process of spontaneous symmetry breaking
this kind of model developes a minimum that violates R-parity.  As
there are no neutral fields carrying baryon number, it will always
remain unviolated. Electrically neutral sneutrinos, however, carry
lepton number, so that a nonvanishing vacuum expectation value (VEV)
of a sneutrino would lead to lepton number violation and breaking of
the R-parity~\cite{lsneutrinovev}. In some versions of SUSYLR model a
non-vanishing sneutrino VEV is in fact
unavoidable~\cite{sneutrinovevunavoidable,Huitu:1999tg}. The R-parity
violating interactions are then determined by the spontaneous symmetry
breaking mechanism.

Much work on R-parity breaking by sneutrino VEVs has been done in the
framework of MSSM with explicit R-parity breaking
terms~\cite{mssmvev}. One of the main differences between R-parity
breaking MSSM and SUSYLR models are that physical spontaneous symmetry
breaking is very non-trivial in SUSYLR models, strongly restricting
the parameters of the model. SUSYLR model has more gaugino and
Higgsino fields than MSSM, and as a result, there is a set of R-parity
violating Yukawa interactions that are unique to the SUSYLR
models. Left-right models give also a solution to the neutrino mass
puzzle: the neutrino mass is naturally small due to so-called see-saw
mechanism. If the R-parity is broken spontaneously the nature of the
see-saw mechanism that gives the neutrino mass changes, as the
neutrino is mixed with Higgsinos.

In~\cite{Huitu:1999tg} the Higgs sector of the left-right models with
spontaneous R-parity violation was studied in detail. In this work I
will study the mass spectrum and couplings of the Higgs fields more in
detail. I will also investigate the fermion sector and the R-parity breaking
couplings in this class of models.

In this work a bottom-up approach will be used: first I define four
phenomenologically viable models having gauged $B-L$ symmetry. I
discuss the Higgs sector of these models. The R-parity breaking
manifests itself in the fact that some scalar and fermion mass
eigenstates are mixtures of fields with different R-quantum numbers. I
give mass formulae and compositions for physical charged and neutral
lepton fields in terms of the model parameters, and analyze their
interactions with Higgs fields and gauge bosons. A summary of
resulting R-parity breaking Yukawa interactions is given. In order to
handle large fermion mass matrices we need to use some approximative
methods, which are described in the appendix.

\section{Description of models}

The minimal left-right models involving gauged $U(1)_{B-L}$ symmetry
can be divided into two classes: either the right-handed symmetry
breaking is accomplished by the VEV of the right-handed
sneutrino VEV (models 1a and 1b), in which case the right-handed scale
is limited to the TeV range; or there are $SU(2)_R$ triplet fields
that contribute to the symmetry breaking (models 2a and 2b). In the
latter case the right-handed scale, and thus mass of the extra gauge
bosons, can be arbitrarily heavy.

By minimal we mean that the models have minimal phenomenologically
acceptable supersymmetric particle content for a chosen gauge symmetry
group and for a chosen scale of vacuum expectation values. We do not,
however, set any {\em a priori} constraints to the couplings of the
model. In the following we list the particle content of four such
models. The spontaneous R-parity violation is unavoidable in three of
these models: in models 1a and 1b a non-vanishing sneutrino VEV is
needed to give phenomenologically acceptable masses ($\sim 1$~TeV) to
the right-handed gauge bosons. In model 2b the R-parity must also be
spontaneuously broken, unless the model is expanded with
non-renormalizable interaction terms or extra Higgs
fields~\cite{sneutrinovevunavoidable,Huitu:1999tg}. Model 2a has both
R-parity violating and conserving physical vacuum solutions. In this
work we, however, concentrate solely on the R-parity violating
solutions.

\subsection{Model~1a: $U(1)_R$ and $v_R \sim 1{\text{TeV}}$}

The minimal SUSYLR model obeying gauge symmetry $SU(2)_L \times U(1)_R
\times U(1)_{B-L} \times SU(3)_C$ has the same chiral superfields as
MSSM, except that there are additional right-handed neutrino
superfields. The superfield content of the model is thus the
following ($i=1,2,3$):
\begin{eqnarray}
L_L^i = \left( \begin{array}{c} \nu_L^i \\ e_L^i \end{array} \right) &
\left( {\bf
2},0,-\frac 12, {\bf 1} \right) ,
\nonumber \\
e_R^i& \left( {\bf
1},\frac  12,\frac 12, {\bf 1} \right)  , \nonumber \\
\nu_R^i & \left( {\bf
1},- \frac  12,\frac 12, {\bf 1} \right)  , \nonumber \\
Q_L^i = \left( \begin{array}{c} u_L^i \\ d_L^i \end{array} \right) &\left( {\bf
2},0,-\frac 16, {\bf 3} \right)  , \nonumber \\
d_R^i &\left( {\bf
1},\frac 12,\frac 16, {\bf 3^*} \right) , \nonumber \\
u_R^i &\left( {\bf
1},-\frac 12,\frac 16, {\bf 3^*} \right)  , \nonumber \\
\phi_{1} = \left( \begin{array}{c} \phi_{11}^0 \\ \phi_{11}^- \end{array} \right) & \left( {\bf
2},-\frac 12,0, {\bf 1} \right)   , \nonumber \\
\phi_2 = \left( \begin{array}{c} \phi_{22}^+ \\ \phi_{22}^0 \end{array} \right) & \left( {\bf
2},\frac 12,0, {\bf 1} \right) .
\end{eqnarray}
The most general renormalizable superpotential for these fields can be
written as
\begin{equation}
W_{1a}=\lambda_\nu \phi_2^T i \tau_2 L_L \nu_R+\lambda_e \phi_1^T i
\tau_2 L_L e_R+\lambda_u \phi_2^T i \tau_2 Q_L u_R+\lambda_d \phi_1^T
i\tau_2 Q_L d_R+\mu_\phi
\phi_1^T i\tau_2 \phi_2 ,
\end{equation}
where generation indices have been suppressed. 
The resulting scalar potential is minimized by the following set of VEVs:
\begin{equation}
\langle \tilde \nu_R \rangle = \sigma_R \simeq v_R, \text{ } \langle
\phi^0_{11} \rangle = v_d, \text{ } \langle \phi^0_{22} \rangle =
v_u , \text{ } \langle \tilde \nu _{Lk} \rangle = \sigma_{Lk}.
\label{eq:1avevs}
\end{equation}
The symmetry breaking proceeds at two stages: at scale $v_R$ $U(1)_R
\times U(1)_{B-L}$ is broken by sneutrino VEV $\sigma_R$ to the
hypercharge symmetry $U(1)_Y$ of the standard model. The residual
$SU(2)_L \times U(1)_Y$ symmetry is further broken to $U(1)_{em}$ at
the weak scale. The gauge couplings of respective symmetry groups obey
relation $g_Y^{-2}(v_R)=g_R^{-2}(v_R)+g_{B-L}^{-2}(v_R)$.

The
$\sigma_R$ appears in D-terms in squark mass-squared matrices. The VEV
$\sigma_R$ is at most of the order of the soft supersymmetry
breaking mass squared terms of the SM quarks ($\tilde m_{Q_L}^2$ and
$m_{Q_R}^2$), if the $U(1)_{em} \times SU(3)_C$ gauge symmetry is to
remain unbroken \cite{Huitu:1997iy}:
\begin{equation}
\frac 18 g_R^2 |D| \alt \frac 12 \left( \tilde m_{Q_L}^2+\tilde
m_{Q_R}^2 \right) \sim
\left(1\text{TeV} \right) ^2 ,
\label{eq:susylimit}
\end{equation}
where $\tilde m_{Q_L}^2$ and $\tilde m_{Q_R}^2$ are the soft mass
squared terms for the squarks and where the D-term is in model 1a
\begin{equation}
D \equiv \sigma_R^2 .
\label{eq:dterm1}
\end{equation}

The sneutrino VEV $\langle \tilde \nu_R \rangle = \sigma_R$
contributes, along with the VEVs of the Higgs doublets, to the mass of
the right-handed gauge bosons $W_R$ and $Z_R$:
\begin{eqnarray}
m_{W_R}^2 = \frac 12 g_R^2 \left( \sigma_R^2+v_d^2+v_u^2
\right) , \nonumber \\
m_{Z_R}^2 \simeq \frac 12 \left(g_R^2+g_{B-L}^2 \right) \left(
\sigma_R^2+v_d^2+v_u^2 \right) .
\label{eq:gaugemass1}
\end{eqnarray}

The VEV of the left-handed sneutrino $\langle \tilde \nu _L \rangle =
\sigma_L$ contributes to the mass of the $W_L$-boson, which is given by
\begin{equation}
m_{W_L}^2 = \frac 12 g_L^2 \left( v_u ^2+ v_d ^2 + \sigma_L^2 \right) .
\label{eq:gaugemass2}
\end{equation}
The physical top quark mass is related to the Yukawa coupling
$\lambda_t$ by the ${\overline{MS}}$ relation $m_t/ (1+4 \alpha_s/3
\pi)=\lambda_t v_u$. If mass of the top quark $m_t$ is taken to be
$m_t = 175$~GeV, the requirement that the Yukawa coupling $\lambda_t$
is in perturbative region ($\lambda_t^2 < 4 \pi$) yields the limit
$\sigma_L \alt 168 \text{ GeV}$. This limit could be further improved
to by requiring that the top Yukawa coupling remains perturbative upto
some higher scale. Requiring perturbativity upto the GUT scale $\sim 2
\times 10^{16}$~GeV sets the limit to about
$\sigma_L\alt 90$~GeV.

\subsection{Model 1b: $SU(2)_R$ and $v_R \sim 1{\text{TeV}}$}

In order to make the parity symmetry explicit the right-handed gauge
group can be promoted from $U(1)_R$ of model 1a to $SU(2)_R$. Explicit
parity symmetry thus motivates one to extend the gauge group to
$SU(2)_L \times SU(2)_R \times U(1)_{B-L} \times SU(3)_C$. As in model
1a, the left-right symmetry group is broken to the MSSM symmetry group
at scale $v_R \agt 1$~TeV.

The chiral superfields of the minimal version of the model 
are ($i=1,2,3$):
\begin{eqnarray}
L_L^i = \left( \begin{array}{c} \nu_L^i \\ e_L^i \end{array} \right) &
\left( {\bf 2},{\bf 1},-\frac 12, {\bf 1} \right) , \nonumber \\ L_R^i
= \left( \begin{array}{c} e_R^i \\ \nu_R^i \end{array} \right) &\left(
{\bf 1},{\bf 2},\frac 12, {\bf 1} \right) , \nonumber \\ Q_L^i =
\left( \begin{array}{c} u_L^i \\ d_L^i \end{array} \right) &\left(
{\bf 2},{\bf 1},-\frac 16, {\bf 3} \right) , \nonumber \\ Q_R^i =
\left( \begin{array}{c} d_R^i \\ u_R^i \end{array} \right) &\left(
{\bf 1},{\bf 2},\frac 16, {\bf 3^*} \right)  ,
\nonumber \\ \phi_k = \left( \begin{array}{cc} \phi^0_{1k} &
\phi^+_{2k} \\ \phi^-_{1k} & \phi^0_{2k} \end{array} \right) & \left(
{\bf 2},{\bf 2},0, {\bf 1} \right) \text{ }(k=1,2) .
\end{eqnarray}
The fields obtain VEVs as in model 1a equation~(\ref{eq:1avevs}). The
equations~(\ref{eq:susylimit}) and~(\ref{eq:dterm1}) for D-term as
well as~(\ref{eq:gaugemass1}) and~(\ref{eq:gaugemass2}) for gauge
boson masses are valid also in this case. As in model 1a, the
sneutrino VEV $\langle \tilde \nu _R \rangle = \sigma_R$ contributes
to the masses of right-handed gauge bosons.  Note that as long as the
right-handed scale $v_R$ is close to the supersymmetry breaking scale
$M_{SUSY}$, as defined by equation~(\ref{eq:susylimit}), the $SU(2)_R$
triplet fields are not needed for symmetry breaking. The
superpotential can be written as
\begin{equation}
W_{1b}= L_L^T i \tau_2 \left( \lambda_\nu \phi_2 +\lambda_e \phi_1
\right) L_R+Q_L^T i \tau_2 \left( \lambda_d \phi_2+\lambda_u \phi_1
\right) Q_R + \sum_{i,j=1}^2 {\mu_\phi^{ij} {\text{Tr}} \phi_i i
\tau_2 \phi^T_j i \tau_2 } ,
\end{equation}
where lepton and quark family indices have been suppressed.

We have checked that there is a realistic radiative symmetry breaking
by explicitly calculating the full physical scalar spectrum. 

\subsection{Model 2a: $U(1)_R$ and $v_R \gg 1 {\text{TeV}}$}

In order to have a physical symmetry breaking the D-terms can be at
most of the supersymmetry breaking scale. All squarks would not have
physical masses, if the D-term related to $U(1)_R$ and
$U(1)_{B-L}$ gauge symmetries would be large (see
equation~(\ref{eq:susylimit})). In order to facilitate the
right-handed symmetry breaking at some large scale $v_R\gg M_{SUSY}$
one must add fields that cancel out the large contributions to the
D-terms.

The minimal anomaly-free addition to model~1a that cancels the large
contributions to the D-term related to the {\em both} $U(1)_R$ {\em
and} $U(1)_{B-L}$ gauge symmetries is a pair of $\delta$-fields:
\begin{eqnarray}
\delta_R & \left( {\bf 1}, -1,1,{\bf 1} \right)  , \nonumber \\
\Delta_R & \left( {\bf 1}, 1,-1,{\bf 1} \right) 
\end{eqnarray}
The most general gauge-invariant renormalizable superpotential is
\begin{equation}
W_{2a}=W_{1a}+f_R \nu_R \nu_R \Delta_R+\mu_{\Delta R} \delta_R
\Delta_R .
\end{equation}

These fields will obtain VEVs $\langle \delta_R \rangle =
v_{\delta_R}$ and $\langle \Delta_R \rangle =v_{\Delta_R}$.  The
D-term related to $U(1)_R$ and $U(1)_{B-L}$ gauge groups is then
\begin{equation}
\left| D \right| \equiv \left| \sigma_R^2+2
v_{\delta_R}^2-2 v_{\Delta_R}^2 \right| \alt M_{SUSY}^2 .
\label{eq:dterm}
\end{equation}
Model 2a has been studied extensively in the case of conserved
R-parity in \cite{2amodel}.

\subsection{Model 2b: $SU(2)_R$ and $v_R\gg 1 {\text{TeV}}$}

In this case, as in model 2a, the D-terms can be at most of the order
of the supersymmetry breaking scale. As before, in order to cancel the
contributions both to the D-term related to the both $SU(2)_R$ and
$U(1)_{B-L}$ gauge symmetries one must introduce extra fields in
addition to those appearing in model~1b. The minimal addition is a
pair of triplet superfields:
\begin{eqnarray}
\Delta_R = \left( \begin{array}{cc} \frac 1 {\sqrt 2} \Delta_R^- &
\Delta_R^0 \\ \Delta_R^{--} & - \frac 1 {\sqrt{2}}  \Delta_R^-
\end{array} \right) & \left( {\bf
1},{\bf 3},- 1, {\bf 1} \right) , \nonumber \\
\delta_R = \left( \begin{array}{cc} \frac 1 {\sqrt{2}} \delta_R^+ &
\delta_R^{++} \\ \delta_R^0 & - \frac 1 {\sqrt{2}}  \delta_R^+
\end{array} \right) & \left( {\bf
1},{\bf 3}, 1, {\bf 1} \right) .
\label{eq:triplets}
\end{eqnarray}
Model 2b has been studied in
\cite{model2bnor,Huitu:1999tg,Huitu:1997iy,model2b,fcnc}. In the
minimum of the scalar potential these fields acquire non-vanishing
VEVs $\langle \Delta_R^0 \rangle = v_{\Delta_R}$ and $\langle
\delta_R^0 \rangle = v_{\delta_R}$. One can, in order to preserve
explicit left-right symmetry, add corresponding $SU(2)_L$ triplet
fields $\Delta_L$ and $\delta_L$. With suitable choice of parameters
they decouple from the scalar and fermion mass matrices. Therefore,
for simplicity, they will not be taken into account in the following
discussion.

The superpotential of the model is
\begin{equation}
W_{2b}=W_{1b}+f_R L_R^T i \tau_2 \Delta_R L_R+\mu_{\Delta R}
{\text{Tr}} \Delta_R \delta_R .
\end{equation}The spontaneous R-parity breaking is unavoidable in this model
\cite{sneutrinovevunavoidable}, the sneutrino having necessarily a
non-vanishing VEV, $\langle \tilde \nu_R \rangle = \sigma_R \ne 0$, in
all minima of the scalar potential that conserve the electric charge.
One could, however, modify the model in such a way that the sneutrino
VEV vanishes and there is no R-parity violation. This could be done,
for example, by adding one $SU(2)_R$ triplet that is singlet under
$U(1)_{B-L}$ or by introducing some non-renormalizable operators to
the superpotential~\cite{Huitu:1999tg,nonren}.

In appendix~\ref{sec:shape} we have found a global minimum for the
models~2a and~2b. At the limit of large right-handed scale $v_R$ the
right-handed VEVs $v_{\delta_R}$, $v_{\Delta_R}$ and $\sigma_R$ are
typically of the same order
\begin{equation}
\sigma_R \sim v_{\delta_R} \sim v_{\Delta_R} \sim v_R ,
\end{equation}
while the D-term~(\ref{eq:dterm}) is of the order of the supersymmetry
breaking scale $M_{SUSY}^2 \sim \left( 1{\text{TeV}} \right) ^2 \ll
v_R^2$. In particular, it would be natural to have the sneutrino VEV
$\sigma_R$ of the order of the right-handed scale $v_R$.

Large VEV $\sigma_R$ takes the model away from the supersymmetric
minimum of the scalar-potential. This could result in need for fine
tuning in model parameters. Fine-tuning is not needed, however, if the
couplings obey the following relations (see
appendix~\ref{sec:stability}):
\begin{equation}
|\mu_\phi| \alt M_{SUSY}, \text{ } |\mu_{\Delta R}| \alt M_{SUSY} ,
\text{ } |\lambda_\nu \sigma_R| \alt M_{SUSY} {\text{ and }} | f_R v_R |
\alt M_{SUSY} .
\label{eq:noqdivergences2}
\end{equation}

\section{Higgs spectrum}

In all models, the scalar sector is larger than that of the MSSM. The
requirement that the minimum of the scalar potential conserves
electric charge and color (i.e. all scalar mass-squared eigenvalues
are non-negative) restricts the parameter space. A numerical example
of full Higgs spectrum of model~2b is given in
appendix~\ref{sec:example}.

\subsection{Light neutral Higgs scalar} 

The light Higgs spectrum is
characterized by one light neutral Higgs scalar
\begin{equation}
h \simeq  \cos \beta
\text{Re} (\phi^0_{11}) + \sin \beta \text{Re} (\phi^0_{22})  ,
\label{eq:lightesthiggs}
\end{equation}
where $\tan \beta=v_u/v_d$ is the ratio of Higgs bidoublet VEVs.  It
has a tree-level upper limit for its mass~\cite{Huitu:1998rr,Huitu:1999tg}:
\begin{equation}
m_{h}^2 \le \left( 1+ \frac {g_R^2}{g_L^2} \right) m_{W_L}^2 \cos ^2 2 \beta .
\label{eq:lighthiggs}
\end{equation}
The radiative corrections to limit~(\ref{eq:lighthiggs}) have been
calculated in~\cite{Huitu:1999tg}, and it was found that they increase
the tree-level upper bound on the mass of $m_h$ typically by $\sim
30$GeV.

The limit~(\ref{eq:lighthiggs}) can be made stricter by taking the
heavy ($\sim m_{Z_R}$) Higgs direction into account. The $2\times 2$
submatrix $M^2$ of the full mass matrix of model 2b is
\begin{eqnarray}
M^2_{11} &= & \frac 12 \left( g_L^2+ g_R^2 \right)
M_L^2 \cos ^2 2 \beta , \nonumber \\
M^2_{12}=M^2_{21}&= & - 2 \lambda_\nu^2 M_L M_R \sin^2 \beta x^2+
\frac 12 g_R^2 M_L M_R (- \cos 2 \beta ) , \nonumber \\
M^2_{22}&= & \frac 12 \left
( g_R^2+g_{B-L}^2 \right) M_R^2+24 f_R \mu_{\Delta R} v_{\delta R}
x^2-12 f_R A_R v_{\Delta R} x^2  -\frac{16 \mu_{\Delta R} B_{\Delta R}
v_{\Delta R} v_{\delta R}}{M_R^2} +4 f_R^2 x^2 \left(\sigma_R^2-8
v_{\Delta R}^2  \right) 			    ,
\end{eqnarray}
where the scalar fields are taken to the light
direction~(\ref{eq:lightesthiggs}) and to the direction $\frac 1N (2
v_{\Delta_R} \text{Re} (\Delta_R^0)-2 v_{\delta_R} \text{Re}
(\delta_R^0)-\sigma_R \text{Re}(\tilde \nu_R ))$ corresponding to the
heavy Higgs, which we will discuss later in this section.  We have
used $M_L^2=v_u^2+v_d^2=2 m_{W_L}^2/g_L^2$, $M_R^2=\sigma_R^2+4 v_{\Delta
R}^2+4 v_{\delta R}^2=2 m_{Z_R}^2/(g_R^2+g_{B-L}^2)$ and
$x=\sigma_R/M_R$. The limit~(\ref{eq:lighthiggs}) can be saturated
only if the non-diagonal element $M_{12}^2$ is small, that is, the
product of neutrino Yukawa coupling $\lambda_\nu$ and R-parity
breaking parameter $x$ is $\lambda_\nu x \sin \beta \sim g_R |\cos 2
\beta|^{1/2}/2$.

Matrix $M^2$ yields an upper limit for the mass of the lighter Higgs
boson (at least if fine-tuning conditions~(\ref{eq:noqdivergences2})
are satisfied):
\begin{eqnarray}
m_h^2 \le & \frac 12 \left( g_L^2 + g_R^2 \right) M_L^2 \cos ^2 2
\beta-\frac{1}{2 \left(g_R^2+g_{B-L}^2 \right)} M_L^2 \cos^2 2 \beta
\left( g_R^2+4 \lambda_\nu^2 \sin^2 \beta x^2/\cos 2 \beta \right) ^2+
{\cal{O}} \left( M_{SUSY}^4/M_R^2 \right) \nonumber \\ & = m_{Z_L}^2
\cos ^2 2 \beta + \frac{1}{g_R^2+g_{B-L}^2 } 4 \lambda_\nu^2 \sin^2
\beta x^2 M_L^2 (-\cos 2 \beta) \left( g_R^2 - 2 \lambda_\nu^2 \sin^2
\beta (-\cos 2 \beta) x^2 \right) + {\cal{O}} \left( M_{SUSY}^4/M_R^2
\right) ,
\label{eq:noqd}
\end{eqnarray}
where we have used $m_{Z_L}^2=\frac 12 (g_L^2+g_Y^2) M_L^2$.  At the
limit of no R-parity breaking ($x=0$) and large right-handed scale
($M_L \ll M_R$) the mass limit reduces to the MSSM result.

\subsection{Triplet Higgs bosons} 

The model~2b contains phenomenologically
very interesting triplet Higgs fields. The masses of $SU(2)_L$
triplets $\Delta_L$ and $\delta_L$ are free parameters of the theory:
at supersymmetric limit their mass is given by the mu-term
$\mu_{\Delta L}$. The masses of $SU(2)_R$ triplet fields $\Delta_R$
and $\delta_R$ are, on the other hand, strongly constrained by the
spontaneous symmetry breaking mechanism. One of the most exciting
predictions specific to the left-right models is the existence of the
doubly charged Higgs fields. The doubly charged Higgs field could be
very light, and they can potentially be seen at LHC or at planned
electron-positron linear collider~\cite{doublycharged}.

Combining equation~(\ref{eq:dterm1}) with results about Higgs boson
mass limits presented in
appendix~\ref{sec:hbml}, equations~(\ref{eq:mlim1})
and~(\ref{eq:mlim2}), one finds
\begin{equation}
4 f_R^2 v_{\Delta_R}^2 \le  f_R A_{f_R} v_{\Delta_R}+ f_R
\mu_{\Delta R} v_{\delta_R} \le 8 f_R^2 \left( v_{\Delta_R}^2-\frac
13 v_{\delta_R}^2 \right) ,
\label{eq:combined}
\end{equation}
where terms of order ${\cal{O}} (M_{SUSY}^2)$ have been ignored. 

The minimization conditions of the scalar potential
\begin{eqnarray}
\frac 1{v_{\Delta_R}} \frac{\partial V}{\partial v_{\Delta_R}} = &2
\mu_{\Delta_R}^2+2 B_{\Delta_R} \mu_{\Delta_R} \frac
{v_{\delta_R}}{v_{\Delta_R}}+4 \left(v_{\Delta_R}^2 -v_{\delta_R}^2
\right) \left( 4 f_R^2-\frac {f_R A_{f_R}}{v_{\Delta_R}} \right)
+{\cal{O}} \left( M_{SUSY}^2 \right) &=0,
 \\
\frac 1{\sigma_R} \frac{\partial V}{\partial \sigma_R} = & -4 f_R
A_{f_R} v_{\Delta R}-4 f_R \mu_{\Delta R} v_{\delta_R}+8 f_R^2 \left(2 
v_{\Delta_R}^2-v_{\delta_R}^2 \right) +{\cal{O}} \left( M_{SUSY}^2
\right) &=0 ,
\label{eq:minc}
\end{eqnarray}
can be realized only if 
\begin{equation}
|\mu_{\Delta R}| \sim | f_R v_{\Delta_R} | \equiv | f_R v_R | 
{\text{ or }} |\mu_{\Delta R}|,| f_R v_R | \alt M_{SUSY}.
\end{equation}

Combining the equations~(\ref{eq:combined}) and~(\ref{eq:minc}) it
follows from the minimization of the potential that
\begin{equation}
|\mu_{\Delta R} | \alt  M_{SUSY}  \text{ and } | f_R
v_{\Delta_R} | = | f_R v_R | \sim M_{SUSY} .
\end{equation}
These conditions are similar to equations~(\ref{eq:noqd}), which where
obtained by requiring no fine-tuning.  Because the mu-term
$\mu_{\Delta R}$ is constrained to be of the order of the
supersymmetry breaking scale, there are only two heavy ($m \gg
M_{SUSY}$) Higgs fields: one neutral scalar field
\begin{equation}
\frac 1N \times
 \left(
 2 v_{\Delta_R}  {\text{Re}} \left( \Delta_R^0 \right)- 
 2 
v_{\delta_R} {\text{Re}} \left( \delta_R^0 \right) - \sigma_R 
{\text{Re}} \left( \tilde \nu_R \right) \right) ,
\label{eq:heavy1}
\end{equation}
with mass $m^2 \simeq \frac 12 (g_R^2+g_{B-L}^2) ( 4 v_{\Delta_R}^2+4
v_{\delta_R}^2+\sigma_R^2 ) \simeq m_{Z_R}^2$ and a charged Higgs
field
\begin{equation}
\frac 1N \times
\left( 2 v_{\Delta_R} v_{\delta_R} \delta_R^\pm - \left( \sigma_R^2+2
v_{\delta_R}^2 \right) \Delta_R^\pm +\sqrt 2 \sigma_R v_{\delta_R}
\tilde e^\pm_R \right) ,
\end{equation}
with mass $m^2 \simeq \frac 12 g_R^2 ( 2 v_{\Delta_R}^2+2
v_{\delta_R}^2+\sigma_R^2 ) \simeq m_{W_R}^2$.

If we would have extended the Higgs sector by $U(1)_{B-L}$ singlet
$SU(2)_R$ triplet or if we would have had some non-renormalizable
operators in the superpotential, we would have in supersymmetric
minimum two Higgs fields at the right-handed scale $v_R$, while most
of the scalar degrees of freedom would have a mass around
$v_R^2/M_{Planck} \ge M_{SUSY}$, with $v_R$ being at least
$10^{10}$GeV in non-renormalizable model~\cite{nonren}.

\subsection{Additional Higgs doublets and CP-violation}

In models~1b and~2b the spectrum of Higgs bosons is quite large. They
have total of four $SU(2)_L$ Higgs doublets. Two of them correspond to
the MSSM doublets related to the electroweak symmetry breaking. The
other two extra Higgs doublets can induce dangerous flavour-changing
neutral currents, that would result in unacceptably large mass
splitting and CP violation for $K^0$, $D^0$ and $B^0$ mesons. The
limits on CP violation can set a lower limit of ${\cal{O}}(20{\text{
TeV}})$ to the mass of the neutral flavour changing Higgs bosons
$\phi^0_{12}$ and $\phi^0_{21}$~\cite{fcnc}.

The CP violating processes can be suppressed by a suitable definition
of left-right symmetry. There are two possible ways to define the
left-right symmetry in terms of the quark Yukawa matrices (see
appendix~\ref{sec:lr}):
\begin{eqnarray}
 \lambda_d=\pm \lambda_d^T , & \lambda_u=\pm
\lambda_u^T {\text{ and }} \label{eq:lr1a} \\
 \lambda_d= e^{i \alpha} \lambda_d^\dagger , & \lambda_u= e^{i
\beta} \lambda_u^\dagger . \label{eq:lr2a}
\end{eqnarray}

The contribution of the mass matrices to the strong CP phase is at
tree level $\text{Arg Det}(M_u M_d)$, where $M_u$ and $M_d$ are mass
matrices for the up- and down-quarks, respectively. This contribution
to the strong CP-phase would automatically vanish, if the Yukawa
matrices are hermitean, as in symmetry defined by
equation~(\ref{eq:lr2a}), and if the vacuum expectation values of the
Higgs bosons are real~\cite{kuchimoha}.

The extra Higgs bosons contribute to flavour changing neutral currents
(FCNC) in $K-\overline K$ mixing. These contributions can set a lower
limit of ${\cal{O}}(20\text{ TeV})$ to the mass of the extra Higgs
bosons~\cite{fcnc}. The contribution to the phase term is proportional
to ${\text{Im}}( (V_L^* D_d V_R^\dagger)_{uc} (V_L^* D_d
V_R^\dagger)_{cu}^* )$, where $V_L$ and $V_R$ are the two
Cabibbo-Kobayashi-Maskawa present in the left-right-models, and $D_d$
is diagonal matrix $D_d={\text{diag}}(m_d,m_s,m_b)$. If the model
obeys symmetry of equation~(\ref{eq:lr1a}) the imaginary phase term
vanishes. In this case the model is invariant under left-right
transformation defined by
\begin{equation}
\phi_k \leftrightarrow -\tau_2 \phi_k^T \tau_2, \text{ } Q_L
\leftrightarrow Q_R , \text{ } L_L \leftrightarrow L_R .
\end{equation}

\section{Composition and mass of leptons}

If sneutrino has a non-vanishing vacuum expectation value then the
Higgs bosons will mix with slepton fields and physical
neutrinos or charged leptons will in general be mixtures of
gauginos, higgsinos and lepton interaction eigenstates. In the
following we will describe the lepton sector in models 2a and
2b. Similar results would apply also for models 1a and 1b.

The mass matrices are quite large. In
appendix~\ref{sec:fermionmasses} we present some approximative methods
to compute the masses and compositions of the lightest charginos and
neutralinos (the physical leptons). In this section we will just
discuss on the results.

\subsection{Model 2a}

The chargino and neutralino mass Lagrangian is of the form (see
e.g. \cite{Haber:1985rc})
\begin{equation}
{\cal{L}}=- \frac 12 \left( \begin{array}{cc} \Psi^{+T} & \Psi^{-T}
\end{array} \right) \left(
\begin{array}{cc} 0 & X^T \\ X & 0 \end{array} \right) \left(
\begin{array}{c} \Psi^{+} \\ \Psi^{-}
\end{array} \right) -\frac 12 \Psi^{0T} Y \Psi^0+ \text{h.c. } .
\end{equation}
In model 2a $\Psi^{+T}=(-i \lambda_L^+,\tilde \phi^+_{22},e^+_R )$,
$\Psi^{-T}=(-i \lambda_L^-,\tilde \phi^-_{11},e^-_L )$ and
\begin{equation}
X = \left(
\begin{array}{ccc} M_L & g_L v_u & 0 \\ g_L v_d & \mu_\phi & \lambda_e
\sigma_L \\ g_L \sigma_L & - \lambda_\nu \sigma_R & - \lambda_e v_d
\end{array} \right) .
\end{equation}
For neutralinos $\Psi^{0T}=(-i \lambda^0_L,-i\lambda^0_R,-i
\lambda^0_{B-L},\tilde \phi^0_{11},\tilde \phi^0_{22},\tilde
\Delta_R^0,\tilde \delta_R^0,\nu_L,\nu_R)$. Upper triangle of
symmetric matrix $Y=Y^T$ is
\begin{equation}
Y=\left( \begin{array}{ccccccccc} M_L & 0 & 0 & \frac 1{\sqrt{2}} g_L
v_d & -\frac 1{\sqrt{2}} g_L v_u & 0 & 0 & \frac 1{\sqrt{2}} g_L
\sigma_L & 0 \\ & M_R & 0 & -\frac 1{\sqrt{2}} g_R v_d & \frac
1{\sqrt{2}} g_R v_u & \sqrt{2} g_R v_{\Delta_R} & - \sqrt{2} g_R v_{\delta_R} & 0 & -\frac 1{\sqrt{2}} g_R \sigma_R \\
& & M_{B-L} & 0 & 0 & -\sqrt{2} g_{B-L} v_{\Delta_R} & \sqrt{2} g_{B-L} v_{\delta_R} & 0 & \frac 1{\sqrt{2}} g_{B-L} \sigma_R \\
& & & 0 & -\mu_\phi & 0 & 0 & 0 & 0 \\
& & &   & 0 & 0 & 0 & \lambda_\nu \sigma_R & \lambda_\nu \sigma_L \\
& & &   &   & 0 & \mu_{\Delta R} & 0 & -2 f_R \sigma_R \\
& & &   &   &   & 0 & 0 & 0 \\
& & &   &   &   &   & 0 & \lambda_\nu v_u \\
& & &   &   &   &   &   & -2 f_R v_{\Delta_R} \end{array} \right) .
\end{equation}

Let us define a dimensionless parameter $\tan \alpha_L$ that describes
the strength of R-parity breaking couplings in model 2a:
\begin{equation}
\tan \alpha_L = \frac{ \lambda_\nu \sigma_R}{\mu_\phi} .
\end{equation}
The composition of physical charged lepton is then
\begin{equation}
\tau = \left( \begin{array}{c} \sin \alpha_L \tilde \phi^-_{11}+\cos
\alpha_L e_L^- \\ \overline{e_R^+} \end{array} \right) ,
\end{equation}
and mass is given by
\begin{equation}
m_\tau = \left| \lambda_e \left( \sin \alpha_L \sigma_L-\cos \alpha_L
v_d \right) \right|
.
\end{equation}
The composition of physical neutrino is
\begin{equation}
\nu_\tau = \sin \alpha_L \tilde \phi^0_{11} + \cos \alpha_L \nu_L ,
\end{equation}
An approximation for neutrino mass can be calculated using methods
described in appendix~\ref{sec:fermionmasses}. Instead of giving the
neutrino mass formula in its complete form we present here the result
at the limit of large right-handed scale ($v_{\Delta_R} \gg
M_{SUSY}$):
\begin{equation}
m_{\nu_\tau} \simeq \left| \frac 12 \left[ \frac{g_L^2}{M_L}    +\left(
\frac{M_{B-L}}{g_{B-L}^2}+\frac{M_R}{g_R^2} \right) ^{-1}  \right] \left( \cos \alpha_L \sigma_L+\sin \alpha_L v_d
\right) ^2  
- \frac{\lambda_\nu^2 v_u^2 \cos^2 \alpha_L}{2 f_R v_{\Delta_R}}
\right| .
\label{eq:neutrinomass}
\end{equation}
Equation~(\ref{eq:neutrinomass}) is a reasonable approximation of the
eigenvalue of the full mass matrix, since the $v_{\Delta_R}$ is
expected to be at least at multi-TeV range. At the limit of no
R-parity breaking ($\sigma_R=\sigma_L=0$) the neutrino mass formula
reduces to the normal see-saw relation $m_{\nu_\tau}=\lambda_\nu^2
v_u^2/(2 f_R v_{\Delta_R})$.  Due to the constraint $f_R v_R \sim
M_{SUSY}$ one would expect the neutrino mass always to be of the order
$m_\tau^2/M_{SUSY}$.

The sneutrino VEVs $\sigma_L$ contribute also to the neutrino
masses. At the limit of vanishing Yukawa couplings
$\lambda_\nu=\lambda_e=0$ and universal gaugino masses
$M=M_L=M_R=M_{B-L}$ the neutrino mass can be approximated by
equation~(\ref{eq:neutrinomass}). Using the current experimental
limits on neutrino masses~\cite{Caso:1998tx},
\begin{equation}
m_{\nu_e}<10\text{eV}, \text{ } m_{\nu_\mu} < 0.17 \text{MeV}, \text{
} m_{\nu_\tau} < 18 \text{MeV} ,
\label{eq:neutrinomasslimits}
\end{equation}
one obtains the following upper limits for the sneutrino VEVs
$\sigma_{Lk}=\langle \tilde
\nu_{Lk} \rangle$:
\begin{equation}
\left| \sigma_{Le} \right| < 0.004\text{GeV} \left( \frac{M}{\text{TeV}}
\right)^{\frac 12} ,
\text{ }
\left| \sigma_{L \mu} \right| < 0.6\text{GeV} \left( \frac{M}{\text{TeV}} \right)^{\frac 12} ,
\text{ }
\left| \sigma_{L \tau} \right| < 6\text{GeV} \left( \frac{M}{\text{TeV}} \right)^{\frac 12} .
\label{eq:sllimits}
\end{equation}
applying to all models discussed in this work.

Taking the limits on neutrino masses~(\ref{eq:neutrinomasslimits})
into account one can constrain the angle $\alpha_L$ for lepton family
(at limit $\sigma_L=0$ and when the gaugino contribution dominates 
the neutrino masses):
\begin{equation}
\left| \sin \alpha_{Le} \right| \alt \frac{7\times 10^{-5}}{\cos
\beta} \left( \frac{M_{gaugino}}{\text{TeV}} \right) ^{\frac
12} , \text{ } \left| \sin \alpha_{L \mu} \right| \alt
\frac{0.009}{\cos \beta} \left( \frac{M_{gaugino}}{\text{TeV}} \right) ^{\frac
12}  , \text{ } \left| \sin \alpha_{L \tau} \right| \alt
\frac{0.1}{\cos \beta} \left( \frac{M_{gaugino}}{\text{TeV}} \right) ^{\frac
12} .
\label{eq:alphalraja}
\end{equation}
For the third lepton family the mixing is unrestricted for large
values of $\tan \beta \agt 10$.\footnote{These constraints could be
relaxed if the gaugino and triplet contributions to the neutrino mass
were tuned to cancel out.}

\subsection{Model 2b}

The main difference between models 2a and 2b comes at the chargino
sector: charged lepton can have components from the $SU(2)_R$ gaugino
field and $SU(2)_R$ triplet Higgsino fields. The composition of the
charged lepton is
\begin{equation}
\tau = \left(  \begin{array}{c} \sin \alpha_L' \cos \alpha_L'' \tilde
\phi^-_{11} + \sin \alpha_L' \sin \alpha_L '' \tilde
\phi^-_{21}+\cos \alpha_L' e_L^- \\ \sin \alpha_R \cos \alpha_R' \overline{(-i
\lambda^+_R)}+  \sin \alpha_R \sin \alpha_R' \overline{\tilde
\delta^+_R}- \cos \alpha_R \overline{e_R^+} \end{array} \right) .
\end{equation}
The angles $\alpha_R$, $\alpha_L'$ and $\alpha_L''$ are defined in
appendix~\ref{sec:angles}.

Due to approximate $SU(2)_L$ symmetry the physical neutrino is similar to the
left-handed part of the physical charged lepton:
\begin{equation}
\nu_\tau = \sin \alpha_L' \cos \alpha_L'' \tilde
\phi^0_{11} + \sin \alpha_L' \sin \alpha_L '' \tilde
\phi^0_{21}+\cos \alpha_L' \nu_L .
\end{equation}

The $SU(2)_R$ gaugino component in the right-handed part of the
physical charged lepton is phenomenologically interesting: a large
gaugino component in the physical lepton will result to lepton-number
violating Yukawa operators that are specific for SUSYLR models. At the
limit of large right-handed scale $v_R\gg M_{SUSY}$ and setting
$v_{\delta_R}=0$ and $M_R=\mu_{\Delta_R}=M_{SUSY}$ one has in leading
order
\begin{equation}
\tan \alpha_R = \left( \frac{g_R \sigma_R} {M_{SUSY}} \right) ^2,
\text{ } \tan \alpha_R' = \frac{ g_R \sigma_R}{M_{SUSY}} .
\end{equation}
At this limit the right-handed part of the physical lepton is composed
mostly of triplet Higgsinos ($\tilde \delta _R^+$) and $SU(2)_R$
gauginos ($-i \lambda_R^+$). The gaugino component in physical lepton
can thus be quite large for moderately large sneutrino VEV $\sigma_R$.

\section{Fermion couplings to bosons}

The physical processes where R-parity violation manifests itself will
most probably include fermions. In this last section we discuss the
Yukawa couplings and anomalous gauge couplings of the quarks and
leptons.

\subsection{Coupling to Higgs boson}

The chargino mass Lagrangian can be written in the form
\begin{equation}
{\cal{L}} = - \Psi^{-T} X \Psi^+ + \text{h.c.} = -\chi^{-T} D
\chi^+ +\text{h.c. } .
\end{equation}
where $D$ is a diagonal positive definite matrix and $\chi^+=V \Psi^+$
and $\chi^-=U \psi^-$ and $X=X_0+X_1$ is the chargino mass
matrix. $X_1$ contains all terms that are proportional to the VEVs
that transform non-trivially under $SU(2)_L$, while $X_0$ contains all
terms proportional to the supersymmetry breaking parameters and
$SU(2)_L$ singlet VEVs.

We define unitary matrices $U_0$ and $V_0$ to be such that $D_0=U_0^*
X_0 V_0^\dagger$ is a diagonal matrix with non-negative entries, and
$(D_0)_{11}=0$ ($X_0$ has in our case one zero eigenvector that corresponds
to the physical lepton mass eigenstate).

At the limit of vanishing anomalous charged lepton coupling to
$Z_L$ boson the mass of the physical lepton is
\begin{equation}
m_1 \approx \left( U_0^* X_1 V_0^\dagger \right)_{11} .
\end{equation}

At decoupling limit the lightest Higgs boson is~\cite{Comelli:1996xg}
\begin{equation}
h = \frac 1v \sum_k {\left( \langle {\text{Re}}\phi_k \rangle
{\text{Re}}\phi_k + \langle {\text{Im}}\phi_k \rangle
{\text{Im}}\phi_k \right) } ,
\end{equation}
where $\phi_k$ are all scalar doublet fields of the theory and
$v^2=\sum_k {\langle \phi_k^* \phi_k\rangle }$. In our case this is
equivalent to equation~(\ref{eq:lightesthiggs}).

A tree-level Lagrangian describing the coupling of the lightest
chargino (the charged physical lepton) $\chi^\pm_1 (\sim \tau)$ to the
Higgs boson $h$ is~\cite{Huitu:1999tg}
\begin{equation}
{\cal{L}}\simeq -\chi^-_1 \left( U_0^* X_1 V_0^\dagger \right) _{11}
\chi^+_1 h + \text{ h.c } = \left(1 +{\cal{O}} \left(\frac {m_W}{M_H}
\right)^2 \right) \frac {m_\tau}v \tau^+ \tau^- h + \text{ h.c. } .
\end{equation}
At decoupling limit chargino coupling thus approaches the standard
model prediction for the Higgs coupling, even if the physical lepton
would be composed mainly of Higgsinos or gauginos.

\subsection{Couplings to the weak currents}
\label{sec:weak}

The lepton mass eigenstates are mixtures of lepton interaction
eigenstates, higgsinos and gauginos. All of these components do not
necessarily have the same $SU(2)_L \times U(1)_Y$ quantum numbers as
the standard model leptons. As a consequence, the lepton couplings to
weak currents are non-universal and different from their standard
model prediction.

The correction term for the neutral current couplings are given in
appendix~\ref{sec:fermionmasses} in equation~(\ref{eq:ncurrcorr}).
The corrections to the axial and vectorial couplings are thus
typically of order $a^2/M_{SUSY}^2$. Since the charged lepton mass is
of the order $m_{\tau} \sim a$ and the neutrino mass is $m_\nu \sim
a^2/M_{SUSY}$, typical perturbation to the axial or vectorial current
would be $\delta A \sim \delta V \sim m_\nu^2/m_\tau^2 \sim
m_\nu/m_{SUSY}$. If the neutrino masses are at their experimental
upper bounds one would expect the perturbations to be
\begin{equation}
 \delta A_e \sim 4
\times 10^{-10}, \text{ } \delta A_\mu \sim 3 \times 10^{-6} , \text{
} \delta A_\tau \sim 1 \times 10^{-4} .
\end{equation}
The experimental resolution is of the order $10^{-3}$. In other words,
the mass limits on neutrinos are generally more restricting than the
limits obtained from the neutral current universality. Only the limit
on tau family can be interesting, if the neutrino mass is close to its
experimental bound and the model parameters are chosen appropriately.

The standard model prediction for the axial current is $A_\tau^{SM}=
-\frac 12$. Assuming that two sigma deviation from the standard model
prediction is acceptable~\cite{Caso:1998tx}, the axial current can
differ from the standard model prediction by
\begin{equation}
\left| \delta A_\tau \right|=\left| A_\tau-A_\tau^{MS} \right| <
0.0026 \text{ } .
\label{eq:atauexp}
\end{equation}

One can derive an analytic expression for the deviation of the axial
and vector
current: 
\begin{equation}
\left. \begin{array}{c} \delta A \\ \delta V \end{array} \right\}
=\delta L\mp \delta R=\frac 12 \left( \sigma_L \cos \alpha_L+ v_d
\sin \alpha_L \right) ^2 \left(-\frac{g_L^2}{M_L^2} \mp \frac {\lambda_\nu^2}
{\mu_\phi^2} \right) .
\end{equation}
When compared to the expression for the neutrino mass
(\ref{eq:neutrinomass}), one sees that the deviation from the standard
model prediction is typically less than
$m_{\nu_\tau}/M_{gaugino}$. The anomalous coupling to weak current is
thus practically always less than the experimental error in the
measurement. (The anomalous coupling can however be large if the ratio
$\lambda_\nu^2/\mu_\phi^2$ is big enough: $|\mu_\phi | \ll | \lambda_\nu
M_L/g_L |$.) 

Similar result applies to charged weak current, since both physical
neutrino and charged lepton mass eigenstates obey $SU(2)_L$ symmetry
to a good accuracy. The $SU(2)_L$ breaking mixing angles in lepton
mass eigenstates are typically suppressed by factor
$\sqrt{m_\nu/M_{SUSY}}$, as shown above for neutral weak current.

\subsection{R-parity breaking couplings}

Most of the R-parity breaking couplings are suppressed either by the
large right-handed scale, by non-observation of heavy neutrinos or by
experimental constraints on the universality of neutral and charged
weak currents. There are, however, a limited set of dimension three
operators that break R-parity and that can be large. All
R-parity breaking Yukawa operators that couple to two standard model
fermions and to a scalar field are listed.

For
simplicity, only one lepton family is taken to have non-vanishing sneutrino
VEV(s). We denote with $k$ the index of this family ($\langle \tilde
\nu _{Rk} \rangle \ne 0$), with $i$ an arbitrary lepton or quark family
and with $j$ an arbitrary lepton or quark family that satisfies $j \ne
k$.

The physical leptons have a Higgsino component. The mixing in model 1a
or 2a is proportional to angle $\alpha_L$. This results into following
effective operators:
\begin{equation}
{\cal{L}}_{2a}=- \lambda_{di} \sin \alpha_L \left( \overline{d^c_i}
P_L \nu_k \tilde d_{Ri}+ \overline {d_i} P_L \nu_k \tilde d
_{Li}-\overline{ u^c_i} P_L e_k \tilde d_{Ri}-\overline {d_i} P_L
e_k \tilde u_{Li} \right)+\text{ h.c. } .
\end{equation}
The lepton-number violating couplings are proportional to the
down-quark Yukawa couplings. All couplings are parametrized by mixing
angle $\tan \alpha_{Lk}$, which is constrained by neutrino masses (see
eq.~(\ref{eq:alphalraja})).

In model 2a one has also Higgsino $\tilde \phi^-_{21}$ components and
$-i\lambda_R^+$ gaugino components mixed in the physical lepton mass
eigenstate. These fields can induce couplings that are proportional to
up-quark Yukawas and gauge coupling $g_R$:
\begin{eqnarray}
{\cal{L}}_{2b}&=&- \sin \alpha_L' \left( \lambda_{di} \cos
\alpha_L''+\lambda_{ui} \sin \alpha_L'' \right) \left(
\overline{d^c_i} P_L \nu_k \tilde d_{Ri}+ \overline {d_i} P_L \nu_k
\tilde d _{Li}-\overline{ u^c_i} P_L e_k \tilde d_{Ri}-\overline {d_i}
P_L e_k \tilde u_{Li} \right) \nonumber \\ && -g_R \sin \alpha_R \cos \alpha_R'
\overline{u^c_i} P_R e_k \tilde d_{Ri}+\text{ h.c. } .
\label{eq:nytjovasyttaa}
\end{eqnarray}
The last term in equation~(\ref{eq:nytjovasyttaa}) is a unique lepton
number violating coupling. It couples universally, with the same
strength, to all (s)quark families. Further, it is not suppressed by
the Yukawa couplings. It is thus the only large R-parity violating
coupling that involves light quark and lepton families. Since the
coupling is due to mixing of $SU(2)_R$ it is also a unique prediction
of R-parity violating SUSYLR models

The R-parity violating operators involving (s)leptons are similar to
those involving quarks. The only difference is that some operators are
cancelled out, if all sleptons involved are from the family having
non-zero sneutrino VEV. The operator proportional to the gauge
coupling involves heavy right-handed neutrino, so it will not be
listed here. The operators are in model 2b are the following:
\begin{eqnarray}
{\cal{L}}&=&- \sin \alpha_L' \left( \lambda_{ej} \cos
\alpha_L''+\lambda_{\nu j} \sin \alpha_L'' \right) \left( \overline{e_j^c}
P_L \nu_k \tilde e_{Rj}+\overline{e_j} P_L \nu_k \tilde
e_{Lj}-\overline {\nu^c_j} P_L e_k \tilde e_{Rj}- \overline {e_j} P_L
e_k \tilde \nu_{L j} \right) \nonumber \\
&& - \sin \alpha_L' \cos \alpha_L ' \left( \lambda_{ek} \cos
\alpha_L''+\lambda_{\nu k} \sin \alpha_L'' \right) \left( \overline
{e_k} P_L \nu_k \tilde e_{Lk}- \overline{e_k} P_L e_k \tilde
\nu_{lk} \right)   .
\label{eq:slept2}
\end{eqnarray}
The result for model 2a is obtained from equation~(\ref{eq:slept2}) by
replacing $\sin \alpha_L' \left( \lambda_{ei} \cos
\alpha_L''+\lambda_{\nu j} \sin \alpha_L'' \right)$ by $\sin \alpha_L
\lambda_{ei}$.

The trilinear R-parity breaking couplings in models 1a and 2a are
similar to those in MSSM with sneutrino VEVs. Models 1b and 2b have
two distinct features:
\begin{itemize}

\item There is proportionality to down {\em and} up quark Yukawa
matrix $\lambda_u$ due to Higgsino components $\tilde \phi_{12}^{0 ,
\pm}$ in the physical leptons.

\item There is a contribution due to the $SU(2)_R$ gaugino in the
right-handed part of the physical lepton
\begin{equation}
{\cal{L}}=-g_R \sin \alpha_R \sin \alpha_R' \tilde d_{Ri}
\overline {u^c_i} P_R e_k +
\text{ h.c. } .
\end{equation}
The contribution due to gaugino is universal for all (s)quark
families. The mixing angles $\alpha_R$ and $\alpha_R'$ are {\em a
priori} free parameters, while the left-handed mixing angles
($\alpha_L$, $\alpha_L'$ and $\alpha_L''$) are constrained by the
neutrino masses~(\ref{eq:alphalraja}).

\end{itemize}

The R-parity breaking vertex proportional to the gauge coupling $g_R$
involves only $SU(2)_L$ singlet fields. The operator could be directly
measured at process $e_k^+ \overline u \rightarrow \tilde d_R^*$ or
$e^- u \rightarrow \tilde d_R$. The latter process could be detected
in HERA, if the electron sneutrino has a non-vanishing VEV
$\sigma_{Re} \ne 0$ and the down-squark $\tilde d_R$ is near the
experimental lower limit on its mass ($\sim 200$~GeV).

A more stringent limit, if the electron sneutrino has a non-vanishing
VEV $\sigma_{Re}\ne 0$, is given by non-observation of neutrinoless
double beta decay. The limit obtained from the lower bound on the
lifetime of $^{76}Ge$ gives~\cite{Faessler:1998zg}
\begin{equation}
\left| g_R \sin \alpha_{Re} \cos \alpha_R' \right| \alt 0.07 \left(
\frac{\tilde m_{d_R}}{\text{TeV}} \right) ^2  \left( \frac{ M_{gluino} 
}{\text{TeV}} \right) ^{\frac 12} ,
\end{equation}
where only the graph involving gluino and to down-squarks $\tilde d_R$ 
has been taken into account.

\section{Conclusion}

I have analyzed a set of minimal models that obey the left-right gauge
symmetries and in which the R-parity is broken spontaneously by a VEV
of a sneutrino. In two of our models (1a and 1b), in which the
right-handed scale is close to the supersymmetry breaking scale, the
$SU(2)_R$ triplet superfields are not needed to have an acceptable
spontaneous symmetry breaking pattern. The VEV of right-handed
neutrino alone is sufficient to make the right-handed gauge bosons
heavy enough. I have analyzed Higgs sectors of these models. The
Higgs sector is characterized by one scalar that at decoupling limit
is like the standard model Higgs boson. The upper limit for its mass
can be
pushed by radiative corrections as high as $150-200$~GeV. In
model 2b at the limit of large right-handed scale there are always
either light doubly charged scalar degree of freedom or a light
neutral singlet degree of freedom.

I have found analytic expressions for masses and mixings of the
neutral and charged leptons. In appendix~\ref{sec:fermionmasses} I
present a general method to calculate the mass eigenvalues and
eigenvectors for large fermion mass matrices. The
experimental bounds on neutrino masses set strict limits on the
left-handed sneutrino VEV and on the anomalous couplings to the
neutral weak current. The deviations to the couplings with the neutral
weak currents are expected to be too small to be observed.

The R-parity breaking trilinear couplings that are
unsuppressed by the low neutrino masses are listed. In model 1a and
2a the lepton number violating trilinear couplings are always
proportional to the mixing angle $\sin \alpha_L$ and the Yukawa
coupling of corresponding quark or lepton family.

In $SU(2)_R$ models the mixing of right-handed part of the charged
lepton with the $SU(2)_R$ gaugino introduces for a universal R-parity
breaking coupling that is proportional to the gauge coupling
$g_R$. This coupling and R-parity breaking coupling proportional to
the up-quark mass matrix can provide unique signature of $SU(2)_R
\times U(1)_{B-L}$ gauge symmetry group.

\appendix

\section{Shape of the potential at right-handed scale}
\label{sec:shape}

The scalar potential of models~1b or~2b expressed in terms of the
right-handed field VEVs can be written as
\begin{eqnarray}
V\simeq \frac 18 \left( g_R^2 +g_{B-L}^2 \right) \left( \sigma_R^2+2
v_{\delta_R}^2-2 v_{\Delta_R}^2 \right) ^2+4 f_R^2 \sigma_R^2
v_{\Delta_R}^2 +\left( \mu_{\Delta R} v_{\delta_R}-f_R \sigma_R^2
\right) ^2 \nonumber +\mu_{\Delta R}^2 v_{\Delta R}^2 \\ +A M_{SUSY}^2 \sigma_R^2+B M_{SUSY}^2
v_{\delta_R}^2+C M_{SUSY}^2 v_{\Delta_R}^2 +D M_{SUSY} \mu_{\Delta R} v_{\delta_R}
v_{\Delta_R}+E f_R M_{SUSY} v_{\delta_R} \sigma_R^2 ,
\end{eqnarray}
where $M_{SUSY}$ is the supersymmetry breaking scale and $A$, $B$,
$C$, $D$ and $E$ are some dimensionless parameters of order unity that
depend on the soft supersymmetry breaking couplings. If we consider a
simplified equation, where we take $\mu_{\Delta R}=D=E=0$, we can
minimize the potential $V$ analytically. There are three possible
solutions for the global minimum of the potential $V$ at the limit of
small $f_R$, corresponding to large right-handed scale $v_R$. The
first solution is the trivial solution
\begin{equation}
\sigma_R=v_{\delta_R}=v_{\Delta_R}=V_{MIN}=0 .
\end{equation}
The second solution is (if $-B-C \ge 0$, $-2A+3B+2C \ge 0$ and
$-2A+2B+C \ge 0$)
\begin{eqnarray}
\sigma_R^2= \frac{-B-C}{4 f_R^2} M_{SUSY}^2, \text{ } v_{\delta_R}^2 =
\frac{-2A+3B+2C}{8 f_R^2} M_{SUSY}^2, \text{ } v_{\Delta_R}^2=
\frac{-2 A+2B+C}{8 f_R^2} M_{SUSY}^2, \nonumber \\
V_{MIN}=- \frac{\left(4A-3B-C \right) \left( B+C \right) }{16 f_R^2}
M_{SUSY}^4 .
\label{eq:sol1}
\end{eqnarray}
and the third solution is (if $-2A-C \ge 0$)
\begin{eqnarray}
\sigma_R^2=2 v_{\Delta_R}^2=\frac{- 2 A-C }{12
f_R^2} M_{SUSY}^2,\text{ } v_{\delta_R}^2=0, \nonumber \\ V_{MIN}=-\frac{ \left(2 A+C \right) ^2 }{48 f_R^2} M_{SUSY}^4.
\label{eq:sol2}
\end{eqnarray}

If, for example, the supersymmetry breaking parameters are chosen to
be $A=-4$, $B=0$ and $C=1$, then solution~(\ref{eq:sol2}) is the global
minimum. The VEVs are then:
\begin{equation}
\sigma_R^2 \simeq 2 v_{\Delta_R}^2 \simeq  \frac 7{12} \frac
{M_{SUSY}^2}{f_R^2} \simeq v_R, \text{ }, v_{\delta_R}^2 \simeq 0 .
\end{equation}
(In the case of model~2b and in this particular case the gauge
couplings should obey $1 < g_{B-L}^2/g_R^2 < \frac {13}7$ for the
global minimum not to break the residual $U(1)_{em}$ gauge symmetry.)

One would thus expect that the right-handed VEVs have the following pattern:
\begin{eqnarray}
\sigma_R^2, v_{\Delta_R}^2, v_{\delta_R}^2 \sim v_R^2 {\text{ or }}
\sigma_R^2,v_{\Delta_R}^2 \sim v_R^2, v_{\delta_R}^2 \sim
{\cal{O}}\left( M_{SUSY}^2 \right) ,
\nonumber \\ 
|D|=|\sigma_R^2+2 v_{\delta_R}^2-2 v_{\Delta_R}^2 |\sim{\cal{O}}\left( M_{SUSY}^2 \right)  .
\label{eq:rpattern2}
\end{eqnarray}

As a result of soft supersymmetry breaking couplings of the order
$M_{SUSY}\simeq {\cal{O}}(1{\text{TeV}})$, it is natural to have the
sneutrino VEV $\sigma_R$ has to be of the order of right-handed scale
$v_R \gg M_{SUSY}$: $\sigma_R \simeq {\cal{O}}(v_R)$. With full mass matrix, taking all parameters into account, that this
is indeed the case.

In the limit of large right-handed scale $v_R \gg M_{SUSY}$ the value
of the potential at minimum is typically quite large: $V_{MIN} \sim
M_{SUSY}^2 v_R^2 \gg M_{SUSY}^4$. One could potentially have large
quadratic corrections to the scalar mass terms. It is shown in
appendix~\ref{sec:stability} that the quadratic correctins are
suppressed if the couplings obey certain relations.

\section{Fine-tuning considerations}
\label{sec:stability}

The quadratic radiative corrections $\delta M^2$ to the scalar masses
are typically of the order
\begin{equation}
\delta M^2 \sim \frac {\lambda^2}{8 \pi ^2} \delta \mu ^2 ,
\end{equation}
where $\lambda$ is some Yukawa or gauge coupling constant and $\delta
\mu ^2$ is typical mass-difference between corresponding scalar and
fermion degrees of freedom. If $\delta \mu^2$ would be large ($\gg
M_{SUSY}^2$) the radiative corrections to the scalar mass terms
$\delta M^2$ could potentially be also large. In other words, one
would have re-introduced the naturalness problem. 

In supersymmetric minimum of model 2a all VEVs vanish. However, we
require some of the VEVs to be much larger than the supersymmetry
breaking scale. This could potentially result in large mass splitting
between fermionic and bosonic degrees of freedom. 

The part of the scalar potential related to the F-terms is
\begin{equation}
V_F= \sum_k \left| F_k \right| ^2 ,
\end{equation}
where $F_k=\partial W/\partial \phi_k$ denotes partial derivative of
superpotential $W$ with respect to a chiral superfield $\phi_k$.  The
contribution of the F-terms to the real scalar mass matrix $\tilde
M_{ij}^2$ is
\begin{equation}
 \sum_k {\frac 12 \left( \frac{\partial^2}{\partial \phi_i \partial \phi_j} -
\frac{\delta_{ij}}{v_i} \frac{\partial}{\partial \phi_i} \right)
\left( \frac{\partial W}{\partial \phi_k} 
\right)^2 } 
%\nonumber \\ 
= \sum_k {\langle \frac{\partial^2 W}{\partial \phi_i \partial \phi_k}
\rangle \langle \frac{\partial^2 W}{\partial \phi_k \partial \phi_j}
\rangle } + \sum_k { \langle F_k \rangle \langle \frac{\partial^2
F_k}{\partial \phi_i \partial \phi_j} - \frac{\delta_{ij}}{v_i}
\frac{\partial F_k}{\partial \phi_i} \rangle } ,
\label{eq:qdiv}
\end{equation}
where $v_i$ denotes the VEV of chiral superfield $\phi_i$. The first
sum in equation~(\ref{eq:qdiv}) gives the supersymmetric mass terms
that are similar to those in the neutralino mass matrix. The part
proportional to F-term $F_k$ contributes to the mass splitting between
scalar and fermion degrees of freedom.

From the scalar mass matrices one can see that the mass difference
$\delta \mu ^2$ in the present model (with a large $\sigma_R$) due to
large $F$-terms is restricted to $\delta \mu ^2 \alt M_{SUSY}^2$,
providing that the model parameters obey the following
relations:\footnote{One could derive the
limits~(\ref{eq:noqdivergences}) also from minimization conditions of
the scalar potential $\partial V/\partial \phi_k=0$ by requiring that
the soft supersymmetry breaking terms are at most of the order of
$M_{SUSY}$. Similar inequalities have been found in the case of Model
2b in~\cite{Huitu:1999tg}.}
\begin{equation}
|\mu_\phi| \alt M_{SUSY}, \text{ } |\mu_{\Delta R}| \alt M_{SUSY} ,
\text{ } |\lambda_\nu \sigma_R| \alt M_{SUSY} {\text{ and }} | f_R v_R |
\alt M_{SUSY} .
\label{eq:noqdivergences}
\end{equation}
Radiative corrections to the scalar potential should thus have no
large quadratic corrections, even if we are in fact quite far from the
supersymmetric minimum of the scalar potential.

Another way to analyze fine tuning is to write the electroweak gauge
boson masses in terms of model parameters at higher scale. In this
case the minimization conditions for the potential yield at tree
level:
\begin{eqnarray}
\frac{ g_R^2}{g_L^2} m_{W_L}^2 \cos 2 \beta= 
m_{\delta_R}^2+ \mu_{\Delta R}^2 +\frac 12 \left( g_R^2+g_{B-L}^2 \right) D+
\frac{ \mu_{\Delta R}}{v_{\delta_R}} \left( v_{\Delta_R} B_{\Delta
R}- f_R \mu_{\Delta_R} \sigma_R^2 \right) , \nonumber \\
\frac{ g_R^2}{g_L^2} m_{W_L}^2 \cos 2 \beta = -
m_{\Delta_R}^2- \mu_{\Delta R}^2  +\frac 12 \left( g_R^2+g_{B-L}^2 \right) D-
4 f_R^2 \sigma_R^2+\frac{- v_{\delta_R} \mu_{\Delta R} B_{\Delta R}+
f_R A_{f_R}
\sigma_R^2}{v_{\Delta_R}}   .
\label{eq:finet}
\end{eqnarray}
If conditions in equation~(\ref{eq:noqdivergences}) apply all terms in
equation~(\ref{eq:finet}) are of order ${\cal{O}}(M_{SUSY}^2)$ and there
is no need for fine-tuned cancellations.

\section{Example model}
\label{sec:example}

\begin{table}
\caption{Physical scalar mass eigenstates for a particular choise of
paremeters in model~2b ($v_{\Delta_R}=10^7$GeV, $v_{\delta_R}=1.2
\times 10^6$GeV, $\sigma_R=1.4 \times 10^7$GeV, $\tan \beta =3$). The $SU(2)_L$ triplet fields $\Delta_L$ and
$\delta_L$ have not been shown. They do not mix with the other scalar fields. Also the
squarks and the first and second family sleptons have been left
out. The model contains light (${\cal{O}} (10 {\text{GeV}} )$) scalar
and pseudoscalar degrees of freedom that are singlets under the
standard model gauge group. There are always necessarily two heavy
scalar degrees of freedom that have a mass of the order of the
right-handed scale. The doubly charged scalar fields have in this
particular case a mass around the supersymmetry breaking scale
$M_{SUSY}$. The mass eigenstates are calculated at {\em tree-level},
the radiative corrections would e.g. increase the mass of the light
Higgs doublet mass eigenstate.}
\begin{tabular}{c|l}
Mass (TeV) &  Composition \\
\tableline            
$1.4 \times 10^4$  & $-0.098 \text{Re} \left( \delta_R^0 \right) + 0.815 {\text{Re}} \left( \Delta_R^0 \right) - 0.572
\text{Re} \left( \tilde \nu_R \right) $ \\
$9.3 \times 10^3$  & $0.085 \delta_R^\pm - 0.707 \Delta_R^\pm + 0.702 e_R^\pm $ \\
$9.6$ & $-0.695 \phi^\pm_{21} - 0.237 \phi^\pm_{22} - 0.639 \phi^\pm_{11} -
0.232 \phi^\pm_{12} $ \\
$9.6$ & $-0.694 \text{Re} \left( \phi^0_{11} \right) - 0.237 \text{Re} \left( \phi^0_{12} \right) +  0.639
\text{Re}\left(  \phi^0_{21} \right) + 0.232 \text{Re} \left( \phi^0_{22} \right) $ \\
$9.6$ & $0.694 \text{Im} \left( \phi^0_{11} \right) + 0.237 \text{Im} \left( \phi^0_{12} \right) + 0.639 \text{Im} \left( \phi^0_{21} \right) + 
   0.231 \text{Im} \left( \phi^0_{22} \right) $ \\
$9.2$ & $0.993 \text{Im} \left( \delta_R^0 \right) + 0.119 \text{Im} \left( \Delta_R^0 \right) $ \\
$9.0$ & $0.993 \text{Re} \left( \delta_R^0 \right) + 0.04 {\text{Re}} \left( \Delta_R^0 \right) - 0.113 \text{Re} \left( \tilde \nu_R \right) $ \\
$8.8$ & $0.993 \delta_R^\pm - 0.12 e_R^\pm $ \\
$8.7$ & $ 0.988 \delta_R^{\pm \pm} - 0.154 \Delta_R^{\pm \pm} $ \\
$6.2$ & $0.646 \text{Re} \left( \phi^0_{11} \right) - 0.291 \text{Re} \left( \phi^0_{12} \right) + 0.672
\text{Re} \left(  \phi^0_{21} \right) -    0.215 \text{Re} \left( \phi^0_{22} \right) $ \\
$6.2$ & $0.646 \phi^\pm_{21} - 0.291 \phi^\pm_{22} - 0.672 \phi^\pm_{11}
+ 0.215 \phi^\pm_{12} $ \\
$6.2$ & $-0.646 \text{Im} \left( \phi^0_{11} \right) + 0.291 \text{Im} \left( \phi^0_{12} \right) + 0.672
\text{Im} \left( \phi^0_{21} \right) - 0.215 \text{Im} \left( \phi^0_{22} \right) $ \\
$3.1$ & $-0.154 \delta_R^{\pm \pm} - 0.988 \Delta_R^{\pm \pm} $ \\
$1.9$ & $0.025 \phi^\pm_{21} + 0.927 \phi^\pm_{22} - 0.375 \phi^\pm_{11}
+  0.008 \phi^\pm_{12} $ \\
$1.9$ & $ 0.025 \text{Re} \left( \phi^0_{11} \right) + 0.927 \text{Re} \left( \phi^0_{12} \right) + 0.375 \text{Re} \phi^0_{21}
-  0.008 \text{Re} \left( \phi^0_{22} \right) $ \\
$1.9$ & $0.025 \text{Im} \left( \phi^0_{11} \right) + 0.927 \text{Im} \left( \phi^0_{12} \right) - 0.375 \text{Im} \left( \phi^0_{21} \right)
+  0.008 \text{Im} \left( \phi^0_{22} \right) $ \\
$1.7$ & $ \tilde \tau _L ^\pm$ \\
$1.7$ & $\text{Re} \left( \tilde \nu _L \right) $ \\
$1.7$ & $\text{Im} \left( \tilde \nu _L \right) $ \\
$0.073$ & $-0.316 \text{Re} \left( \phi^0_{11} \right) - 0.949 \text{Re} \left( \phi^0_{22} \right) $ \\
$0.039$ & $0.07 \text{Re} \left( \delta_R^0 \right) + 0.579 {\text{Re}} \left( \Delta_R^0 \right) + 0.813 \text{Re} \left( \tilde \nu_R \right) $ \\
$0.009$ & $-0.068 \text{Im} \left( \delta_R^0 \right) + 0.568 \text{Im} \left( \Delta_R^0 \right) + 0.82 \text{Im} \left( \tilde \nu_R \right) $
\end{tabular}
\end{table}

\section{Higgs boson mass limits}
\label{sec:hbml}

The mass of the lightest neutral flavour changing Higgs boson,
composed of $\phi^0_{12}$ and $\phi^0_{21}$, is bound by
\begin{eqnarray}
{\cal{O}}(1-10{\text{TeV}})^2 < M_{\phi^0_{12},\phi^0_{21}}^2 \le &
-\cos^2 2 \beta m_{W_L}^2- \frac 12 g_R^2 \cos 2
\beta \left(D+\frac{2 m_{W_L}^2 \cos 2 \beta}{g_L^2} \right) \nonumber
\\ & +  \sigma_R^2 \left( (\lambda_e)^2 \cos^2 \beta-(\lambda_\nu)^2 \sin^2
\beta 
\right) ,
\end{eqnarray}
where $D$ is given in equation~(\ref{eq:dterm1}) or~(\ref{eq:dterm})
for the models 1b and 2b, respectively. One sees that one must either
have a positive D-term, $ \frac 12 g_R^2 D >
{\cal{O}}(1-10{\text{TeV}})^2$, or alternatively $\lambda_e \cos \beta
\sigma_R$ should have a value at least of the order of TeV (if
$\lambda_e$ is tau Yukawa coupling $\sigma_R$ should be larger that
about $100$~TeV). Added together, in model~1b the sneutrino VEV has a
lower limit $\sigma_R \agt 3{\text{ TeV}}$ (or equivalently $m_{W_R}
\agt 1{\text{ TeV}}$).

It turns out that {\em all} bidoublet Higgs bosons can have a mass of
at most of the order $M_{SUSY}$: only parameter that could make them
heavier would be the parameters $\mu_\phi^{ij}$. However, the
minimization conditions for bidoublet fields read (ignoring all
terms of order $M_{SUSY}$)
\begin{equation}
\frac 1{v_u} \frac{\partial V}{\partial v_u}=16 \left( \mu_\phi^{11} \right) ^2
+16 \left( \mu_\phi^{12} \right) ^2\sim M_{SUSY}^2, \text{ } \frac
1{v_d} \frac{\partial V}{\partial v_u}=16 \left( \mu_\phi^{22} \right) ^2 +16
\left( \mu_\phi^{12} \right) ^2 \sim M_{SUSY}^2 .  
\end{equation}
It follows that the mu-terms, and consequently bidoublet Higgs masses,
are also at most of the SUSY-breaking scale.\footnote{The flavour
changing Higgs doublets could be made to have a mass around the
right-handed scale $v_R$ by introducing non-renormalizable operator
$1/M_* {\text{Tr}}
\phi_1 i \tau_2 \phi_2^T i \tau_2 {\text{Tr}} \Delta_R \delta_R$ to
the superpotential~\cite{model2bnor}.}

One can derive the following upper bound to the mass-squared term of a
neutral Higgs scalar from $3\times 3$ submatrix of the full mass
matrix of models~2a and~2b that involves the fields $\Delta_R^0$,
$\delta_R^0$ and $\tilde \nu_R$:
\begin{equation}
0<\left( v_{\Delta_R}^2+v_{\delta_R}^2+\sigma_R^2 \right)
M_{\Delta_R^0,\delta_R^0\tilde \nu_R}^2 \le  \frac 12 \left(
g_R^2+g_{B-L}^2 \right) D^2+4 f_R^2 \sigma_R^2 \left( \sigma_R^2+4
v_{\Delta_R}^2 \right) -3 f_R A_{f_R} \sigma_R^2 v_{\Delta_R} -3f_R
\mu_{\Delta R} v_{\delta_R} \sigma_R^2.
\label{eq:mlim1}
\end{equation}
The doubly charged Higgs boson of the model 2b has the following upper
bound on its mass:
\begin{equation}
0 < \left( v_{\Delta_R}^2+v_{\delta_R}^2 \right)
M_{\Delta_R^{\pm \pm},\delta_R^{\pm \pm}}^2 \le  g_R^2 \left(
v_{\Delta_R}^2-v_{\delta_R}^2 \right) D+ f_R A_{f_R}
\sigma_R^2v_{\Delta_R}-4 f_R^2 \sigma_R^2 v_{\Delta_R}^2 +f_R
\mu_{\Delta R} v_{\delta_R} \sigma_R^2 .
\label{eq:mlim2}
\end{equation}

\section{Most general discrete left-right symmetry}
\label{sec:lr}

The superfield content of models 2a and 2b are explicitly
left-right-symmetric, if also the left-handed triplets $\Delta_L$ and
$\delta_L$ are taken into account. The quark fields and bidoublet
Higgs fields transform in left-right transformations as follows:
\begin{equation}
Q_L \rightarrow U_L Q_L, \text{ } \tilde Q_L \rightarrow U_L \tilde
Q_L , Q_R \rightarrow U_R Q_R, \text{ } \tilde Q_R \rightarrow U_R \tilde
Q_R , \text{ } \phi_i \rightarrow U_L \phi_i U_R^\dagger , \text{ } \tilde
\phi_i \rightarrow U_L \tilde \phi_i U_R^\dagger,
\end{equation}
where the charge-conjugated fields have been used
\begin{equation}
\tilde Q_{L,R}=i \tau_2 Q_{L,R}^* \text{ and } \tilde \phi_i = -i\tau_2
\phi^*_i i \tau_2 ,
\end{equation}
and the left-right transformation is defined as
\begin{equation}
U_{L,R}=\exp \left( -\frac 12 i \epsilon_{L,R}^k \tau_k  \right) .
\end{equation}

The discrete ${\cal Z}_2$ left-right transformation means that the
model, including the quark mass term, remains invariant under
interchange of $U_L$ and $U_R$, and that two consequent left-right
transformations reduce to identity:
\begin{equation}
{\cal{L}}=- \lambda_i^{ab} Q_{La}^T i \tau_2 \phi_i
Q_{Rb}-\lambda_i^{ab*} \tilde Q_{La}^T i \tau_2 \tilde \phi_i \tilde
Q_{Rb} 
\label{eq:masst}
\end{equation}
Clearly as the gauge operators $U_{L,R}$ are swapped $U_L
\leftrightarrow U_R$ the fields must transform as follows:
\begin{eqnarray}
Q_L^a \rightarrow U_{1b}^a Q_R^b+ V_{1b}^a \tilde Q_R^b , \nonumber \\
Q_R^a \rightarrow U_{2b}^a Q_L^b+V_{2b}^a \tilde Q_L^b , \nonumber \\
\phi_i \rightarrow X_{i}^j i \tau_2 \phi_j^T i \tau_2+Y_{i}^j i \tau_2
\tilde \phi_j^T i \tau_2 .
\end{eqnarray}
Since there are no charge-conjugate fields in superpotential, one must
have either $U_i=X=0$ or $V_i=Y=0$. By suitable redefinition of the
quark field $Q_L$ the one can set $U=1$ or $V=1$. Matrix $X$ or $Y$
can in principle be any unitary $2\times 2$ matrix that satisfy $X^2=1$ or
$Y Y^*=1$. Only cases where matrices $X$ and $Y$ are
diagonal will be considered.

There are thus two ways to define the left-right-symmetry in terms of
quark Yukawa matrices:
\begin{eqnarray}
(a) \text{ } V_i=Y=0: & \lambda_d=X_d^d \lambda_d^T,
 & \lambda_u=X_u^u \lambda_u^T ;
\nonumber \\ (b) \text{ } U_i=X=0: & \lambda_d=
Y_d^d \lambda_d^\dagger , & \lambda_u= Y_u^u
\lambda_u^\dagger ; \label{eq:lr2}
\end{eqnarray}
where $X_u^u,X_d^d=\pm 1$ and $|Y_u^u|,|Y_d^d|=1$ is an arbitrary
phase.  
If the Lagrangian of the model, including gauge the couplings and
triplet Higgs fields, obey these left-right symmetries, the symmetry
is also preserved in the renormalization group running of the model.

\section{Mass eigenvalues and eigenvectors for fermions}
\label{sec:fermionmasses}

The chargino and neutralino mass matrices are typically quite large
and cannot be solved analytically. The fermion mass matrix is
generally of the form 
\begin{equation}
{\cal{L}} = - \frac 12 \Psi^T Y \Psi+\text{h.c.}=- \frac 12 \chi^T D
\chi+\text{h.c. } ,
\end{equation}
where $\Psi$ is a vector of Weyl spinors and $Y=Y^T$ is a symmetric
mass matrix. $D$ is a diagonal mass matrix with non-negative entries
and $\chi=N \Psi$ are the fermion mass eigenstates. The unitary matrix
$N$ satisfies $N^* Y N^\dagger=D$, or $D^2=N Y^\dagger Y N^\dagger$.

For Dirac fermions the mass matrix $Y$ is of the form
\begin{equation}
Y= \left( \begin{array}{cc} 0 & X^T \\ X & 0 \end{array} \right) .
\label{eq:dfermionmass}
\end{equation}
The diagonalizing matrix $N$ is
\begin{equation}
N = \left( \begin{array}{cc} V & -U \\ V & U \end{array} \right) ,
\end{equation}
where $V$ and $U$ are unitary matrices such that $D_D=U^* X V^\dagger$
is a diagonal matrix with non-negative entries (see
e.g.~\cite{Haber:1985rc} for further discussion). The eigenvectors of
the Dirac mass matrix come always in pairs having opposite mass
eigenvalues. 
Although the derivation in this section is given for Majorana spinors,
the generalization to Dirac spinors (i.e. charginos) is
straightforward. 

In our case the mass matrix $Y$ can always be decomposed into two
parts $Y=Y_0+Y_1$, where $Y_0$ contains all supersymmetry breaking
terms and all terms that are proportional to vacuum expectation values
that are singlet under $SU(2)_L \times U(1)_Y$. $Y_0^\dagger Y_0$ is
thus always constructed of block-diagonal submatrices of constant
hypercharge $Y^{HC}$.  $Y_1$ contains all terms that are proportional
to VEVs that break the standard model gauge group (in our models
$v_u$, $v_d$ and $\sigma_L$).

In all our cases the matrix $Y_0$ has at least one zero eigenvalue
that approximately corresponds to the physical lepton. The mass of the
lepton is induced by the (small) terms in matrix $Y_1$. Our idea is
first to transform to basis where zero eigenvectors of matrix $Y_0$
are unit vectors. It is enough for purposes of this work to assume
that $Y_0$ has only one zero eigenvalue. In the end I give a general
result for arbitrary number of zero eigenvectors of $Y_0$. First we
transform to basis where the physical lepton eigenvectors are
approximately unit vectors $\tilde v_0^T=(1,0,...,0)$. To this end an
unitary matrix $\hat N_0$ is defined that satisfies\footnote{To find matrix $\hat
N_0$ we need to find the zero eigenvector of $Y_0^\dagger Y_0$. One
can always find an analytical expression for inverse of an arbitrary
matrix. $n$ zero eigenvectors of matrix $Y_0$ can be found from the
basis spanned by $\lim_{\varepsilon \rightarrow 0} \varepsilon ^n
\left( Y_0+\varepsilon {\bf 1} \right) ^{-1}$. }
\begin{equation}
\tilde Y_0=\left( \hat N_0^* Y_0 \hat N_0^\dagger \right)_{ij} =  0,
\text{ } i=1 \text{ or } j=1 .
\end{equation}
We further define matrices $\tilde Y_1=\hat N_0^* Y_1 \hat
N_0^\dagger$, $\tilde Y=\hat N_0^* Y \hat N_0^\dagger$, $a_i=(\hat
N_0^* Y_1 \hat N_0^\dagger)_{1i}$ and $\hat Y_0=( \tilde Y_0 )_{\hat 1
\hat 1}$, where $A_{\hat i \hat j}$ denotes matrix $A$ with row $i$
and column $j$ removed.

We develop the mass of the lightest eigenvector into series with
respect to the eigenvalues of $Y_1$. There are many ways to do it ---
simplest expression is obtained by using determinants. The lepton mass
$m$ is 
\begin{equation}
m= \frac{\left| D \right|}{\left| D_{\hat 1 \hat 1} \right|}=\frac
{\left| N^* \left( Y_0+Y_1 \right) N^\dagger \right| }{\left|\left(
N^* \left( Y_0+Y_1 \right) N^\dagger \right) _{\hat 1 \hat 1} \right|
} = \left\{ \begin{array}{lc} a_1+ {\cal{O}} \left( Y_1^2 \right)
 & , \text{ }  a_1 \ne 0, \\
\sum_{i,j \ne 1} { (-1)^{i+j} a_i a_j \frac{ \left| \left( \hat Y_0
\right) _{\hat i-1, \hat j-1} \right| }{\left| \hat Y_0 \right| }}
+{\cal{O}} \left( Y_1^3 \right) & , \text{ } a_1=0 . \end{array}
\right. 
\label{eq:massformula}
\end{equation}
The ratio
of derivatives in equation~(\ref{eq:massformula}) is simplified by the
fact that $\hat Y_0$ is a block-diagonal matrix. If the blocks are
small enough the ratios reduce to quite simple expressions.

It turns out that the first term $a_1$ dominates the charged lepton
masses. For neutrinos $a_1$ vanishes and the masses are determined to
the leading order by the generalized see-saw formula given by the
sum-term in equation~(\ref{eq:massformula}).

In the mass formula one can essentially approximate the lepton
eigenvector by the zero eigenvector of matrix $\tilde Y_0$. The zero
eigenvector of matrix $\tilde Y_0$ is $\tilde v_0=(1,0, ... ,0)$. To
estimate the accuracy of this approximation and to calculate anomalous
couplings to the weak currents one should know the leading order
corrections to vector $\tilde v_0$: $\tilde v_1=\tilde v_0 +\delta
\tilde v$. The lepton mass is the smallest eigenvalue of the fermion
mass matrix. A standard (numerical) method to find accurate expression
for the smallest eigenvector of matrix is to multiply the
approximation by inverse of the matrix. It is easily seen that this
way the errors of the approximation are reduced at least by factor of
$m/M$, where $m$ is the smallest eigenvalue (physical lepton mass) and
$M$ is the second-smallest eigenvalue of the mass matrix (typically
the lightest supersymmetric chargino or neutralino). 

Thus the leading order correction to vector $\tilde v_0$ is obtained
multiplying it by matrix $\tilde Y ^{-1}$ and normalizing it.
\begin{equation}
\tilde v_1 = \frac{ \tilde Y ^{-1} \tilde v_0 }{ \left| \tilde Y ^{-1}
\tilde v_0 \right| } .
\end{equation}
$\tilde v_1$ can be calculated to leading order
\begin{equation}
\left( \tilde Y ^{-1} v_0 \right)_i = \left( \tilde Y ^{-1}
\right)_{i1}= (-1)^{i+1}\frac{\left| \tilde Y_{\hat i \hat 1} \right| }{ \left|
\tilde Y \right| }\approx \left\{ \begin{array}{lc} \frac 1m & , \text{ } i=1 , \\
\sum_{j \ne 1}{ (-1)^{i+j} a_j \frac{ \left| \left( \hat Y_0
\right) _{\hat i-1, \hat j-1} \right| }{m \left| \hat Y_0 \right| } } &
, \text{ } i \ne 1 .
		      \end{array} \right. 
\end{equation}

The correction $\delta v$ to eigenvector $\tilde v_0$ is thus to the
leading order ($\tilde v_1=\tilde v_0 + \delta \tilde v$)
\begin{equation}
\delta \tilde v_i= \left\{ \begin{array}{cc} 0 & , i=1 \\ \sum_{j \ne 1}{ (-1)^{i+j} a_j \frac{ \left| \left( \hat Y_0
\right) _{\hat i-1, \hat j-1} \right| }{ \left| \hat Y_0 \right| }} &
i \ne 1  		   \end{array} \right. .
\end{equation}

We need expression for $\sum_{i,i' \ne 1} {\delta \tilde v_i C_{ii'}
\delta \tilde v_{i'}'}$ to calculate the anomalous coupling to weak
currents (see section \ref{sec:weak}). Dimension of $\delta \tilde
v_i$ is $N$ and dimension of $\delta \tilde
v_i'$ is $N'$. We can take $N \ge N'$ without loss of generality. We
further assume that we have permuted the basis so that $C$ is a
diagonal matrix with equal diagonal elements grouped together.

Since we want to do algebra with determinants, it is useful to expand
some of the matrices to square form:
\begin{equation}
\hat Y_0'' = \left( \begin{array}{cc} \hat Y_0' & 0_{N'-1 \times
N-N'} \\ 0_{N-N' \times N'-1} & 1_{N-N' \times N-N'} \end{array}
\right) , \text{ } C'=\left( \begin{array}{cc} \hat C_{\hat 1 \hat 1}
&  \begin{array}{c} 0_{N'-1 \times N-N'} \\  1_{N-N' \times N-N'}
\end{array}  	     \end{array} \right) . 
\end{equation}

The required expression is then to leading order
\begin{equation}
\delta \tilde v_i C_{ii'}\delta \tilde v_i' = \sum_{i,i',j,j'} {
(-1)^{i+i'+j+j'} a_j a_{j'}' C_{ii'} \frac{ \left| \left( \hat Y_0 \right)
_{\hat i-1,\hat j-1} \right| \left| \left( \hat Y_0' \right) _{\hat
i'-1,\hat j'-1} \right| }{\left| \left( \hat Y_0 \right) \right|
\left| \left( \hat Y_0' \right) \right| } } = \sum_{j,j'} (-1)^{j+j'}
a_j a_{j'}' C_{ii} \frac{ \left| \left( \hat Y_0^\dagger \hat Y_0 ''
\right)_{\hat j-1,\hat j'-1} \right| }{ \left| \hat Y_0^\dagger 
\hat Y_0 ''  \right| }  .
\label{eq:anomalousweak}
\end{equation}
Due to properties of weak current the matrices appearing in
determinants ($\left( \hat Y_0^\dagger \hat Y_0 ''
\right)_{\hat j-1,\hat j'-1}$ and $\hat Y_0^\dagger 
\hat Y_0 ''$) are block-diagonal matrices with each block
corresponding to a constant $C_{ii}$. The ratios of determinants in
equation~(\ref{eq:anomalousweak}) thus reduces to a ratios of these
diagonal block matrices.

The general neutral weak current coupling for charginos
is of form
\begin{equation}
{\cal{L}}_{NC} = -\frac {g_L}{\cos \theta_W} Z_L^\mu \overline
\Psi_i \gamma_\mu \frac 12 \left( V_i-A_i \gamma^5 \right) \Psi_i =-\frac
{g_L}{\cos \theta_W} Z_L^\mu \overline \Psi_i \gamma_\mu \left( L_i
P_L+R_i P_R \right) \Psi_i ,
\end{equation}
where $P_L=\frac 12 (1-\gamma^5)$ and $P_R= \frac 12
(1+\gamma^5)$. 

The chargino mass matrix is of form~(\ref{eq:dfermionmass}). The
calculation for Dirac fermions proceeds analogously to the Majorana
case discussed above: We define unitary matrices $\tilde U_0$ and
$\tilde V_0$ such that the first row and column of matrix $\tilde
X_0=\tilde U_0^* X_0 \tilde V_0^\dagger$ vanishes. The $a$-vectors are
in this case $a_{Li}=(\tilde U_0^* X_1 \tilde V_0^\dagger ) _{1i}$ and
$a_{Ri}=(\tilde U_0^* X_1 \tilde V_0^\dagger ) _{i1}$.

The correction to the couplings $L$ and $R$ is (as compared to the
standard model value)
\begin{eqnarray}
\delta L = \sum_i { \left( I_{3L}^i-I_{3L}^{e_L} \right) v_i^2 }=
\sum_{i,j,j'} { (-1)^{j+j'} \left( I_{3L}^i-I_{3L}^{e_L} \right)
a_{Lj} a_{Lj'} \frac{ \left| \left( \hat X_0 \hat X_0 ^\dagger \right) _{\hat j
-1,\hat j'-1} \right| }{\left|  \hat X_0 \hat X_0 ^\dagger  \right| }},
\nonumber \\
\delta R = \sum_{i,j,j'} { (-1)^{j+j'} \left( I_{3L}^i-I_{3L}^{e_R} \right)
a_{Rj} a_{Rj'} \frac{ \left| \left( \hat X_0 ^\dagger \hat X_0 \right) _{\hat j
-1,\hat j'-1} \right| }{\left|  \hat X_0 ^\dagger \hat X_0  \right| }} ,
\label{eq:ncurrcorr} 
\end{eqnarray}
where $I_{3L}^i$ are the $SU(2)_L$ quantum numbers for the
corresponding fields (for lepton interaction eigenstates
$I_{3L}^{e_L}=-\frac 12$ and $I_{3L}^{e_R}=0$).

The correction to axial coupling is $\delta A=\delta L-\delta R$ and
to vectorial coupling $\delta V=\delta L+\delta R$.

\section{Fermion mass matrices in model 2b}
\label{sec:angles}

The chargino vectors and mass matrices in model 2b are
\begin{eqnarray}
\Psi^{+T}= & \left( -i \lambda_L^+,-i \lambda_R^+,\tilde \phi^+_{12},\tilde
\phi^+_{22},\tilde \delta^+_R,e^+_R \right),\nonumber \\
\Psi^{-T}= & \left( -i \lambda_L^-,-i \lambda_R^-,\tilde \phi^-_{11},\tilde
\phi^-_{21},\tilde \Delta^-_R,e^-_L \right) , \nonumber \\
X= & \left( \begin{array}{cccccc} M_L & 0 & 0 & g_L v_u & 0 & 0 \\ 0 &
M_R & -g_R v_d & 0 & \sqrt{2} g_R v_{\delta_R} & g_R \sigma_R \\ g_L
v_d & 0 & 2 \mu_\phi^{11} & \mu_\phi^{12} & 0 & \lambda_e \sigma_L \\
0 & -g_R v_u & \mu_\phi^{21} & 2 \mu_\phi^{22} & 0 & \lambda_\nu
\sigma_L \\ 0 & -g_R v_{\Delta_R} & 0 & 0 & \mu_{\Delta_R} & -
\sqrt{2} f_R \sigma_R \\ g_L \sigma_L & 0 & - \lambda_e \sigma_R & -
\lambda_\nu \sigma_R & 0 & -\lambda_e v_d \end{array} \right) 
\end{eqnarray}
For neutralinos $\Psi^{0T}= ( -i \lambda^0_L,-i\lambda^0_R,-i
\lambda^0_{B-L},\tilde \phi^0_{11},\tilde \phi^0_{12},\tilde \phi^0_{21},\tilde \phi^0_{22},\tilde
\Delta_R^0,\tilde \delta_R^0,\nu_L,\nu_R )$ and
\begin{equation}
Y=  \left( \begin{array}{ccccccccccc} M_L & 0 & 0 & \frac 1{\sqrt{2}}
g_L v_d & 0 & 0 & -\frac 1{\sqrt{2}} g_L v_d & 0 & 0 & \frac
1{\sqrt{2}} g_L \sigma_L & 0 \\
& M_R & 0 & -\frac 1{\sqrt{2}} g_R v_d & 0 & 0 & \frac 1{\sqrt{2}} g_R
v_u & \sqrt{2} g_R v_{\Delta_R} & -\sqrt{2} g_R v_{\delta_R} & 0 &
-\frac 1{\sqrt{2}} g_R \sigma_R \\
 & & M_{B-L} & 0 & 0 & 0 & 0 & -\sqrt{2} g_{B-L} v_{\Delta_R} &
\sqrt{2} g_{B-L} v_{\delta_R} & - \frac 1{\sqrt{2}} g_{B-L} \sigma_L &
\frac 1 {\sqrt{2}} g_{B-L} \sigma_R \\
 & &         & 0 & -2 \mu_\phi^{11} & 0 & - \mu_\phi^{12} & 0 & 0 & 0
& 0 \\
 & & & & 0 & - \mu_\phi^{12} & 0 & 0 & 0 & \lambda_e \sigma_R &
\lambda_e \sigma_L \\
 & & & & & 0 & -2 \mu_\phi^{22} & 0 & 0 & 0 & 0 \\
 & & & & & & 0 & 0 & 0 & \lambda_\nu \sigma_R & \lambda_\nu \sigma_L \\
 & & & & & & & 0 & \mu_{\Delta R} & 0 & -2 f_R \sigma_R \\
 & & & & & & & & 0 & 0 & 0 \\
 & & & & & & & & & 0 & \lambda_\nu v_u \\
 & & & & & & & & &  & -2 f_R v_{\Delta_R} \end{array} \right) .
\end{equation}

The angles realted to the right-handed part of the Dirac spinor are
\begin{equation}
\tan  \alpha_R= \sigma_R \frac{\sqrt{ g_R^2 \left( \mu_{\Delta R}+2 f_R
v_{\delta_R} \right) ^2 +2 \left( f_R M_R-g_R^2 v_{\Delta_R} \right)
^2 }}{ M_R \mu_{\Delta_R}+2 g_R^2 v_{\Delta_R} v_{\delta_R}
 } , \text{ } \tan \alpha_R'=\frac {\sqrt 2}{g_R} \frac{f_R M_R-g_R^2 v_{\Delta_R}}{\mu_{\Delta R}+2 f_R
v_{\delta_R}} ,
\end{equation}
and
the angles related to the left-handed part of the Dirac spinor are
\begin{equation}
\tan \alpha_L' = \sigma_R \frac{
\sqrt{\left( \lambda_\nu \mu_\phi^{12}-2 \lambda_e \mu_\phi^{22}
\right)^2+\left( \lambda_e \mu_\phi^{12}-2 \lambda_\nu \mu_\phi^{11}
\right)^2}}{(\mu_\phi^{12})^2-4 \mu_\phi^{11} \mu_\phi^{22}} , \text{ }
\tan \alpha_L'' = \frac{\lambda_e \mu_\phi^{12}-2 \lambda_\nu 
\mu_\phi^{11}}{\lambda_\nu \mu_\phi^{12}-2 \lambda_e \mu_\phi^{22}} .
\end{equation}

\end{document}